\documentclass[epj]{webofc}
\usepackage[varg]{txfonts}
\usepackage{psfrag}
\usepackage{graphicx}
\usepackage{amsmath}

\def\slashchar#1{\setbox0=\hbox{$#1$}
   \dimen0=\wd0
   \setbox1=\hbox{/} \dimen1=\wd1
   \ifdim\dimen0>\dimen1
      \rlap{\hbox to \dimen0{\hfil/\hfil}}
      #1
   \else
      \rlap{\hbox to \dimen1{\hfil$#1$\hfil}}
      /
   \fi}

\def\bei{\begin{itemize}}
\def\ei{\end{itemize}}

\def\beeq{\begin{eqnarray}} 
\def\beqa{\begin{eqnarray}}
\def\bea{\begin{eqnarray}}

\def\eea{\end{eqnarray}}
\def\eqa{\end{eqnarray}}
\def\eeeq{\end{eqnarray}}

\def\eqar{\end{array}}
\def\beqar{\begin{array}}

\def\beas{\begin{eqnarray*}}
\def\beqas{\begin{eqnarray*}}

\def\eqas{\end{eqnarray*}}
\def\eeas{\end{eqnarray*}}

\def\beq{\begin{equation}} 
\def\be{\begin{equation}}

\def\ee{\end{equation}}
\def\eq{\end{equation}}
\def\eeq{\end{equation}}

\def\beqd{\begin{displaymath}}
\def\eeqd{\end{displaymath}}
\def\eqd{\end{displaymath}}

\def\beeq{\begin{eqnarray}} \def\eeeq{\end{eqnarray}}

\newcommand{\fin}{\end{document}}

\newcommand{\veckone}{{\bf k}_1}
\newcommand{\vecktwo}{{\bf k}_2}

\newcommand{\veckjone}{{\bf k}_{J,1}}
\newcommand{\veckjtwo}{{\bf k}_{J,2}}

\newcommand{\deins}[1]{{\rm d}#1\,}
\newcommand{\dzwei}[1]{{\rm d}^2#1\,}

\newcommand{\dkone}{\dzwei{\veckone}}
\newcommand{\dktwo}{\dzwei{\vecktwo}}

\newcommand{\dsigma}{\deins{\sigma}}
\newcommand{\dsigmahat}{\deins{{\hat\sigma}_{\rm{ab}}}}

\newcommand{\dxone}{\deins{x_1}}
\newcommand{\dxtwo}{\deins{x_2}}
\newcommand{\dyjetone}{\deins{y_{J,1}}}
\newcommand{\dyjettwo}{\deins{y_{J,2}}}

\newcommand{\dphijone}{\deins{\phi_{J,1}}}
\newcommand{\dphijtwo}{\deins{\phi_{J,2}}}
\newcommand{\dtwojets}{{\rm d}|\veckjone|\,{\rm d}|\veckjtwo|\,\dyjetone \dyjettwo}

\newcommand{\shat}{{\hat s}}

\newcommand{\init}{\text{init}}

\newcommand{\BLM}{\text{BLM}}

\newcommand{\avgcos}{\langle \cos \varphi \rangle}
\newcommand{\avgcostwo}{\langle \cos 2 \varphi \rangle}

\newcommand{\kina}{$\sqrt{s}=7$ TeV, $|\veckjone|=|\veckjtwo|=35$ GeV}
\newcommand{\kinb}{$\sqrt{s}=14$ TeV, $|\veckjone|=|\veckjtwo|=35$ GeV}
\newcommand{\kinc}{$\sqrt{s}=14$ TeV, $|\veckjone|=|\veckjtwo|=20$ GeV}
\newcommand{\kind}{$\sqrt{s}=14$ TeV, $|\veckjone|=|\veckjtwo|=10$ GeV}

\woctitle{Physics Opportunities at an Electron-Ion Collider}

\begin{document}
\title{High-energy resummation effects in the production of Mueller-Navelet dijet  at the LHC
}

\author{B. Duclou\'e\inst{1,2}\fnsep\thanks{\email{bertrand.b.ducloue@jyu.fi}}\and
        L. Szymanowski\inst{3}\fnsep\thanks{\email{Lech.Szymanowski@ncbj.gov.pl}} \and
        S. Wallon\inst{4,5}\fnsep\thanks{\email{Samuel.Wallon@th.u-psud.fr}}
}

\institute{Department of Physics, University of Jyv\"{a}skyl\"{a}, P.O. Box 35, 40014 University of Jyv\"{a}skyl\"{a}, Finland
\and
Helsinki Institute of Physics, P.O. Box 64, 00014 University of Helsinki, Finland
\and
National Centre for Nuclear Research (NCBJ), Warsaw, Poland
\and
Laboratoire de Physique Th\'eorique, UMR 8627, CNRS, Univ. Paris Sud, Universit\'e Paris-Saclay, \\ \mbox{}\hspace{.03cm} 91405 Orsay, France
\and
UPMC Univ. Paris 06, Facult\'e de Physique,75252 Paris, France }

\abstract{
We study the production of two forward jets with a large interval of rapidity at hadron colliders, which was proposed by Mueller and Navelet as a possible test of the high energy dynamics of QCD, within a complete next-to-leading logarithm framework. We show that using the Brodsky-Lepage-Mackenzie procedure to fix the renormalization scale leads to a very good description of the recent CMS data at the LHC for the azimuthal correlations of the jets. We show 
that the inclusion of next-to-leading order corrections to the jet vertex significantly reduces the importance of energy-momentum non-conservation which is inherent to the BFKL approach, for an asymmetric jet configuration. Finally, we argue that the 
double parton scattering contribution is negligible in the kinematics of actual CMS measurements.
}

\maketitle

\section{Introduction}

 The  dijet production with large rapidity separation, as proposed by Mueller and Navelet~\cite{Mueller:1986ey}, is one of the most promising observables in order to reveal the high energy dynamics of
QCD, described by the Balitsky-Fadin-Kuraev-Lipatov (BFKL)
approach~\cite{Fadin:1975cbKuraev:1976geKuraev:1977fsBalitsky:1978ic}. We here report on our study of this process in a next-to-leading logarithmic (NLL) BFKL resummation, which includes the NLL corrections both to the Green's function~\cite{Fadin:1998py,Ciafaloni:1998gs} and to
the jet vertex~\cite{Bartels:2001ge,Bartels:2002yj}. It has demonstrated that the NLL corrections to the jet vertex have a very large effect, leading to a lower cross section and a much larger azimuthal correlation~\cite{Colferai:2010wu}. These findings are however very dependent on the choice
of the scales, especially the renormalization scale $\mu_R$ and the
factorization scale $\mu_F$, a fact which remains true when using realistic kinematical cuts for LHC experiments~\cite{Ducloue:2013hia}. This dependency can be reduced after using the Brodsky, Lepage and Mackenzie (BLM) scheme~\cite{Brodsky:1982gc}, leading to a very satisfactory description~\cite{Ducloue:2013bva}
of the most recent LHC data extracted by the CMS collaboration for the azimuthal correlations of these jets~\cite{Khachatryan:2016udy}. Our results are not affected,  neither by 
energy-momentum conservation issues nor by potential contribution of multiparton interaction (MPI).

\section{BFKL approach}

The production of two jets of transverse momenta $\veckjone$, $\veckjtwo$ and rapidities
$y_{J,1}$, $y_{J,2}$ is described by the differential cross-section 
\beqa
\label{collinear}
  \frac{\dsigma}{\dtwojets} &=& \sum_{{\rm a},{\rm b}} \int_0^1 \dxone \int_0^1 \dxtwo f_{\rm a}(x_1) f_{\rm b}(x_2)  \frac{\dsigmahat}{\dtwojets},
\eqa
where $f_{\rm a, b}$ are the usual collinear partonic distributions (PDF). In the BFKL framework, the partonic cross-section reads
\beqa
  \frac{\dsigmahat}{\dtwojets}  = \int \dphijone\dphijtwo\int\dkone\dktwo V_{\rm a}(-\veckone,x_1)\,G(\veckone,\vecktwo,\shat)\,V_{\rm b}(\vecktwo,x_2),\label{eq:bfklpartonic}
\eqa
where $V_{\rm a, b}$ and $G$ are respectively the jet vertices and the BFKL Green's function. 
Besides the cross section, the azimuthal correlation of the two jets is another relevant observable sensitive to resummation effects~\cite{DelDuca:1993mn,Stirling:1994he}. 
Defining
the relative azimuthal angle $\varphi$ such that $\varphi=0$ corresponds
to the back-to-back configuration, 
the moments of the distribution 
$  \langle\cos(n\varphi)\rangle$
are of special interest.
Even at NLL accuracy, these observables depend strongly on the choice of the scales, and in particular the renormalization scale $\mu_R$. An
optimization procedure to fix the renormalization scale allows to reduce
this dependency. We use the BLM procedure~\cite{Brodsky:1982gc}, here adapted to the context of BFKL~\cite{Brodsky:1996sgBrodsky:1997sdBrodsky:1998knBrodsky:2002ka}, which is a way of absorbing the non conformal
terms of the perturbative series in a redefinition of the coupling constant, to
improve the convergence of the perturbative series.

\section{Results: symmetric configuration and CMS data}

We compare our results with the  CMS data on the azimuthal correlations of Mueller-Navelet jets at the LHC at a center of mass energy $\sqrt{s}=7$ TeV~\cite{Khachatryan:2016udy}.
The two jets have transverse momenta larger than $35$ GeV and rapidities $0<y_1<4.7$ and $-4.7<y_2<0.$
\begin{figure}[htbp]
\psfrag{Y}[][][1]{\scalebox{.9}{$Y$}}
\psfrag{BLM}[l][l][.7]{\scalebox{.9}{NLL, $\mu_R=\mu_{R,\BLM}$}}
\psfrag{NLL}[l][l][.7]{\scalebox{.9}{NLL, $\mu_R=\mu_{R,\init}$}}
\psfrag{CMS}[l][l][.7]{\scalebox{.9}{CMS data}}
\psfrag{dist}[][][1]{$\frac{1}{\sigma}\frac{d \sigma}{d\varphi}$}
\psfrag{phi}[][][1]{\scalebox{.9}{$\varphi$}}
\hspace{0cm}    \centerline{\includegraphics[height=5cm]{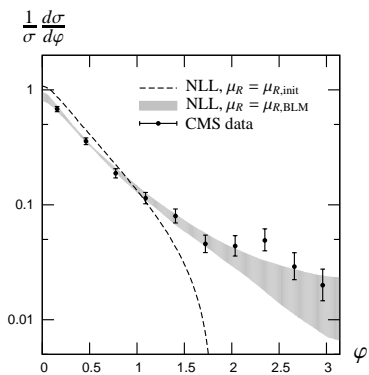}}
  \caption{Symmetric configuration. Azimuthal distribution at NLL accuracy compared with CMS data.}
\label{Fig:cos2cos-dist_blm_sym}
\end{figure}
The improvement due to the BLM procedure is most clearly seen through 
 the azimuthal distribution of the jets 
\begin{equation}
 \frac{1}{{\sigma}}\frac{d{\sigma}}{d \varphi}
  ~=~ \frac{1}{2\pi}
  \left\{1+2 \sum_{n=1}^\infty \cos{\left(n \varphi\right)}
  \left<\cos{\left( n \varphi \right)}\right>\right\}\,,
\end{equation}
as 
displayed in fig.~\ref{Fig:cos2cos-dist_blm_sym}.
On the plot we show the CMS data (black dots with error bars), the NLL BFKL result using the ``natural'' scale choice $\mu_R=\sqrt{|\veckjone| \cdot |\veckjtwo|}$ (dashed black line) and the NLL BFKL results using the BLM scale setting (gray error band). 
We refer to Refs.\cite{Caporale:2012ih,Caporale:2013uva,Caporale:2014gpa,Caporale:2015uva,Celiberto:2015yba,Celiberto:2016ygs} for similar studies.

It is well known that fixed order calculations are unstable when the lower cut on the transverse momenta of both jets is the same, which is the situation encountered 
 by the above CMS measurement~\cite{Khachatryan:2016udy}.
This is due to the fact that in the very peculiar situation where the two jets are almost back-to-back,  resummation effects \`a la Sudakov should be considered, to stabilize the calculation. In the BFKL approach, although this back-to-back limit is stable, the azimuthal distribution can be significantly affected by such resummation effects. These have been obtained recently in the LL approximation~\cite{Mueller:2015ael}. However, this is not available neither at NLL BFKL nor in the fixed order approach. To evade this instability, it would thus be very useful to measure dijet production in a slightly asymmetric configuration, for which the ratio $\avgcostwo/\avgcos$ is expected~\cite{Ducloue:2013bva} to sizeably differ between NLL 
BFKL and next-to-leading fixed order (NLO)~\cite{Aurenche:2008dn} approaches.

\section{Energy-momentum conservation}

Energy-momentum conservation
is not satisfied in the BFKL approach,
being formally a sub-leading effect.
\psfrag{LO}[l][l][.8]{LL}
\psfrag{NLO}[l][l][.8]{NLL}
\psfrag{yeffy}[l][l][.8]{$Y_{\rm eff}/Y$}
\psfrag{kj1}[l][l][0.8]{\hspace{0.1cm}$\veckjtwo$ (GeV)}
\begin{figure}[h]
\vspace{-.2cm}

\centering\includegraphics[width=7cm]{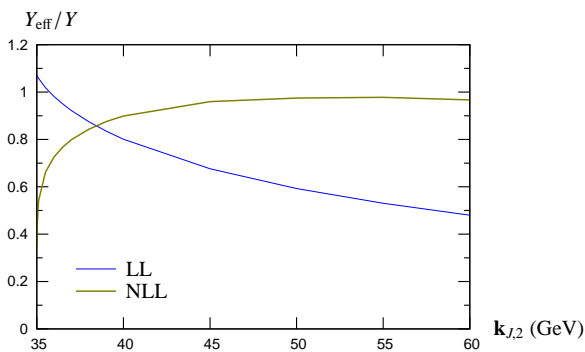}
\caption{Variation of $Y_{\rm eff}/Y$ as defined in eq.~(\protect\ref{eq:Cm_e-m_cons})
as a function of $\veckjtwo$ at fixed $\veckjone=35$ GeV for $Y=8$ and $\sqrt{s}=7$ TeV at leading logarithmic (blue) and next-to-leading logarithmic (brown) accuracy.}
\label{Fig:yeff_NLO}
\end{figure}
It was proposed~\cite{DelDuca:1994ng} to evaluate the importance of this effect by comparing the results of an exact $\mathcal{O}(\alpha_s^3)$ calculation with the BFKL result, expanded in powers of $\alpha_s$ and truncated to order $\alpha_s^3$. 
Having in mind that adding corrections beyond the LL approximation should reduce the violation of
energy-momentum conservation, we here also include NLL corrections to the jet vertices~\cite{Ducloue:2014koa}.
Consider the effective rapidity $Y_{\rm eff}$~\cite{DelDuca:1994ng}
\begin{equation}
  Y_{\rm eff} \equiv\ Y \frac{\mathcal{C}_m^{2\to3}}{\mathcal{C}_m^{{\rm BFKL},\mathcal{O}(\alpha_s^3)}} \,,
  \label{eq:Cm_e-m_cons}
\end{equation}
where $\mathcal{C}_m^{2\to3}$ is the exact $\mathcal{O}(\alpha_s^3)$ result obtained by studying the reaction $gg \to ggg$, while ${\mathcal{C}_m^{{\rm BFKL},\mathcal{O}(\alpha_s^3)}}$ is the BFKL result expanded in powers of $\alpha_s$ and truncated to order $\mathcal{O}(\alpha_s^3)$.
The value of $Y_{\rm eff}$ indicates how valid the BFKL approximation is: a value close to $Y$ means that this approximation is valid, whereas a value significantly different from $Y$ means that it is a too strong assumption in the kinematics under study.
On fig.~\ref{Fig:yeff_NLO} we show the values obtained for $Y_{\rm eff}/Y$ as a function of $\veckjtwo$ for fixed $\veckjone=35$ GeV at a center of mass energy $\sqrt{s}=7$ TeV and for a rapidity separation $Y=8$, in the LL and NLL approximation. As found in ref.~\cite{DelDuca:1994ng}, the LL calculation strongly overestimates the cross section for transverse momenta of the jets not too close, while
at NLL accuracy, the situation is much improved for
significantly different
jet transverse momenta, where instabilities are not expected, see above.

\section{Double parton scattering contribution to MN jets}

\begin{figure}[h]
	\vspace{.1cm}
	
		\centering\includegraphics[height=3.8cm]{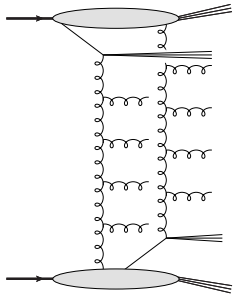}
		\vspace{-.1cm}
		
		\caption{The DPS contribution.}
		\label{Fig:DPS}
	\end{figure}

At high energies and low transverse momenta where BFKL effects are expected to be enhanced, parton densities can become large enough that contributions where several partons from the same incoming hadron take part in the interaction could become important.
We restrict ourselves to the case of double parton scattering where there are at most two subscatterings and where both these scatterings are hard, as illustrated in fig.~\ref{Fig:DPS}.
For simplicity, the order of magnitude of this contribution is evaluated at LL, which we compare with our prediction involving single parton scattering in the BFKL LL and NLL approaches. 
We use a simple factorized ansatz to compute the DPS contribution according to
	\begin{equation}
		\sigma_{\rm DPS}=\frac{\sigma_{\rm fwd} \sigma_{\rm bwd}}{\sigma_{\rm eff}} \, ,
		\label{eq:sigma_DPS}
	\end{equation}
	where $\sigma_{\rm fwd (bwd)}$ is the inclusive cross section for one jet in the forward (backward) direction and $\sigma_{\rm eff}$ is a phenomenological quantity related to the density of the proton in the transverse plane. 
	We vary $\sigma_{\rm eff}$ between 10 and 20 mb, to be consistent with
	the measurements at the Tevatron~\cite{Abe:1993rvAbe:1997xk,Abazov:2009gc,Abazov:2014fha} and at the LHC~\cite{Aad:2013bjm,Chatrchyan:2013xxa}.
\begin{figure}[htbp]
	\vspace{-3.3cm}
	
	\hspace{.9cm}
	\raisebox{0cm}{\includegraphics[width=8.3cm]{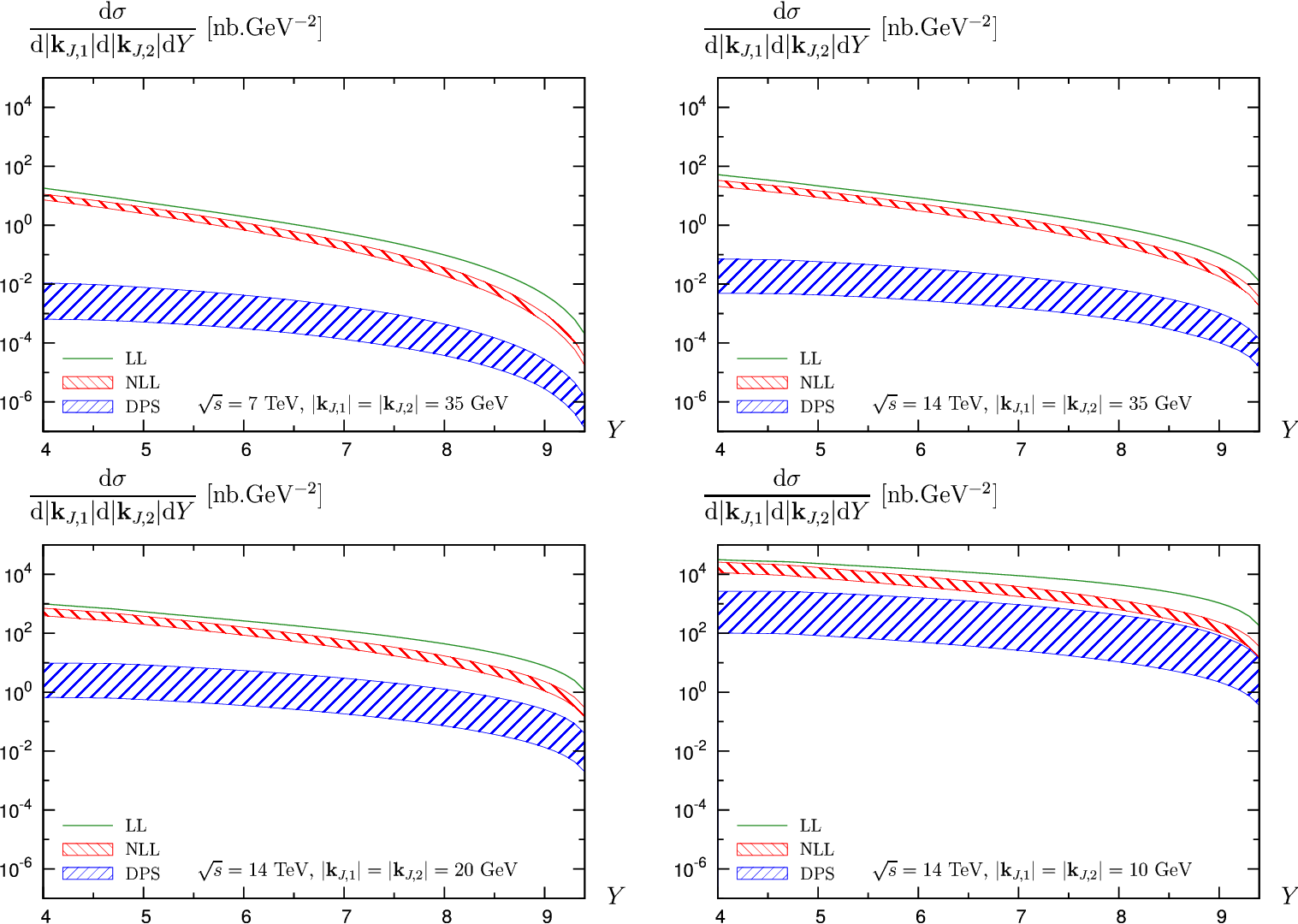}}
	\caption{Comparison of the differential cross section obtained at LL (green) and NLL (red) accuracy in the BFKL approach and the DPS cross section (blue) for the four kinematical cuts described in the text.}
	\label{Fig:sigma}
\end{figure}
The inclusive cross section for one jet in the forward  or backward direction is built as the convolution of the leading order jet vertex with  unintegrated gluon distributions (UGD), 
the global normalization being fitted with CMS~\cite{Chatrchyan:2012gwa} data (see ref.~\cite{Ducloue:2015jba} for more details).
We focus on four choices of kinematical cuts:
\vspace{.2cm}

	 (a) \kina, \phantom{\kinc} \hspace{-4cm} (b) \kinb\,,
	 
	 (c) \kinc, \phantom{\kina} \hspace{-4cm} (d) \kind\,.
\vspace{.2cm}	 
	 
\noindent
The first choice is similar to the cuts used by the CMS analysis of azimuthal correlations of Mueller-Navelet jets at the LHC~\cite{Khachatryan:2016udy}. The other three choices correspond to the higher center of mass energy that the LHC is expected to reach soon. The last two choices correspond to lower transverse momenta at which measurements could become possible in the future, and are particularly relevant since MPI are expected to become more and more important at lower transverse momenta. The rapidities of the jets are restricted according to $0<y_{J,1}<4.7$ and $-4.7<y_{J,2}<0$. We use the MSTW 2008 parametrization~\cite{Martin:2009iq} for collinear parton densities. To estimate the 
uncertainty associated with the choice of the UGD parametrization needed to compute the DPS cross section, we use four different parametrizations~\cite{Kimber:2001sc,Hansson:2003xz,Kutak:2012rf,Hautmann:2013tba}. 
As shown in fig.~\ref{Fig:sigma},
the DPS cross section is always smaller than the SPS one in the LHC kinematics we considered here.
The same conclusion can be reached 
for the impact of double parton scattering on the angular correlation between the jets~\cite{Ducloue:2015jba}.
It is only for the set of parameters giving the largest DPS contribution, {\it i.e.} at low transverse momenta and large rapidity separations, that the effect of DPS can become larger than the uncertainty on the NLL BFKL calculation.

\section{Conclusions}

The azimuthal correlations of Mueller-Navelet jets recently extracted by the CMS collaboration can be well described by a full NLL BFKL calculation supplemented by the use of the BLM  renormalization
scale fixing procedure.
We have shown that two effects which are claimed to have a potential significant impact in this picture do not affect our results. First, the effect of the absence of strict energy-momentum conservation in a BFKL calculation is 
expected to be tiny at NLL accuracy 
 for significantly different values of transverse momenta of the tagged jets. Second, the order of magnitude of DPS contributions is negligible for the kinematics which is under consideration at the LHC. 

\section*{Acknowledgments}

B.~Duclou\'{e} acknowledges support from the Academy of Finland, Project No. 273464. This work was done using computing resources from CSC -- IT Center for Science in Espoo, Finland. L.~Szymanowski was partially
supported by a French Government Scholarship and by a grant of National Science Center, Poland, No. 2015/17/B/ST2/01838.
This work is partially supported by the
French Grant ANR PARTONS No. ANR-12-MONU- 
0008-01.

\end{document}